\documentstyle[12pt,aps]{revtex}

\input{epsf}
\begin{document}

\title{A simple method for monitoring uniformity  of epitaxial semiconductor structures}

\author{G.G.Kozlov\ddag, Yu.K.Dolgikh\ddag, Yu.P.Efimov\dag,
S.A.Eliseev\ddag, \\ V.V.Ovsyankin\ddag and V.V.Petrov\dag}

\maketitle
\vskip20pt
\dag \hskip20pt
Fock Institute of Physics (Petrodvorets Branch), St.Petersburg State University,
Petrodvorets, 198904 Russia

\ddag \hskip20pt Vavilov State Optical Institute, All-Russia Research Center; Birzhevaya
liniya 12, St.Petersburg, 199034 Russia
\vskip20pt
\hskip100pt {\it e}-mail:  gkozlov@photonics.phys.spbu.ru
\vskip20pt

\begin{abstract}
A simple method for visualization of nonuniformity of planar MBE structures is
proposed. The method is based on measuring the relief of the photo-EMF. The
method can be applied to a wide variety of semiconductor structures and does not
require any expensive equipment.
\end{abstract}
\vskip10mm

 The molecular beam epitaxy (MBE) technique allows one to obtain semiconductor structures
 with a specified profile of the bandgap along the growth direction with a spatial
 resolution of atomic range. These potentialities of the MBE technique make it efficient
 in creation of new materials with controllable physical properties. The
 most popular structures obtained, at present, using the MBE technique are quantum wells,
 quantum superlattices, and quantum dots. Ideal structures of the quantum-well and
 superlattice - type, by  definition, should be uniform in the plane of the
 structure.
 The objects like ensembles of quantum dots should be uniform within the scale exceeding
 the mean distance between the dots. Thus, it can be stated that all the above objects
 should be, ideally, macroscopically uniform in the plane of the structure. In reality,
 however, one can never obtain a perfectly uniform planar structure. Nonuniformity of
 the structure is revealed as a dependence of its physical properties (e.g., of spectral
 position of its optical resonance) over the plane of the structure. In some
 cases, this dependence appears to be helpful for experimentalists. It allows
 one, in particular, to  find such a spot on the sample where the spectral position
 of the optical resonance corresponds to the wavelength of the laser used. It should be
 emphasized that the nonuniformities of this kind can be conveniently used when they have
 a smooth monotonic character. In many cases, however, the nonuniformities exhibit a
 complicated spatial dependence which can be hardly employed for practical purposes.
 These nonuniformities are usually of macroscopic scale ($\sim$ 100 $\mu$m) and indicate a
 nonuniformity of the deposition process over the same scale. To detect nonuniformities
 of this scale, it is desirable to have a simple method that enables to plot a map of the
 nonuniformities and to study how they depend on the  growth regime, the type of the
 structure, etc. In [1], there has been proposed a method that allowed one to visualize
 the dependence of the exciton resonance frequency on position on the sample. This method
 allows one to evaluate, in an express way, spectroscopic nonuniformity of the sample. It
 requires, however, a tunable laser and low temperatures. In addition, it is inapplicable
 to samples with no optical resonances. In this note, we describe a simple method of
 mapping the relief of the nonuniformity, which is universally applicable to a great
 variety of semiconductor structures and does not require any expensive equipment. The
 method is based on detection of dependence of the photo-EMF, generated by a
 semiconductor sample, on position of the illuminated spot. In what follows, this
 dependence is referred to as the EMF relief. If the sample is uniform, this dependence is
 absent, while its appearance indicates a nonuniformity of the sample. The magnitude  of the
 spatially dependent component of the EMF is taken, in this method, for the measure of
 the nonuniformity. Schematic of the setup for observation of the photo-EMF relief is
 shown in Fig.1. The laser beam passes through a simple deflecting system comprised of a
 mirror attached to a loudspeaker membrane, and is focused and directed to the sample
 placed in a  capacitor with a transparent upper plate. Illumination of a small spot
 on the surface of the sample produces an electric field in the vicinity of the spot (due
 to the Dember effect) or changes the build-in field. The laser light intensity is
 modulated at a frequency of 15-18 kHz, and the induced field oscillates   with the same
 frequency. As a result, an ac voltage arises on the plates of the capacitor, with
 its amplitude proportional to the photo-EMF generated in the illuminated spot. When the
 sample is nonuniform, the photo-EMF will vary, thus reflecting the nonuniformity of the
 sample. As seen from Fig.1, the laser beam in our setup is swept in two directions as
 it is usually being done in TV monitors. The frame sweep is implemented by a
 slow
 translation of the sample in the direction shown in Fig.1 by a broad arrow. The voltage
 of the capacity varying at the frequency of the laser light modulation is detected and
 stored in a computer. The 2D array thus obtained which represents a relief of the
 photoresponse amplitude over the surface of the sample, is displayed on the monitor as a
 relief of brightness. When the sample under study is uniform, the obtained pattern
 represents a bright area in the form of the sample. Arising   dark and bright spots,
 in this region, indicate a  nonunifornity of the structure. We used, in our setup, a
 commercial laser pointer, with the intensity modulated, at a frequency of 15-18 kHz, by
 modulating the feeding current. The line sweep frequency was 17 Hz. The frame sweep rate
 was chosen to provide 200 lines per frame. As samples for testing, we used epitaxial
 GaAs/AlGaAs structures. Fig.2a shows a relief of the photo-EMF obtained with a pure
 GaAs substrate which we used as a reference. A considerable number of structures grown in
 our laboratory showed a nonuniformity comparable in value with that shown in Fig.2a. To
 illustrate our technique, we present the results of scanning of the samples with
 anomalously strong nonuniformity. Fig.2(b,c) shows the relief of the photo-EMF of an
 epitaxially grown optical waveguide (b) and the same for a GaAs/AlGaAs quantum well (c).
 As seen from these figures the character of nonuniformity of the structures grown on
 the same MBE machine may essentially differ. Note that the nonuniformities shown in Fig.2(b,c)
  cannot be observed using a microscope. At present, we cannot make any definite
 conclusion about correlation (if any) between the nonuniformity of the photo-EMF and
 spectroscopic nonuniformity of the sample. This is the objective of our future studies.
 The study was supported by the International Science and Technology Center (project
 2679)

\newpage
\begin{figure}
\epsfxsize=500pt
\epsffile{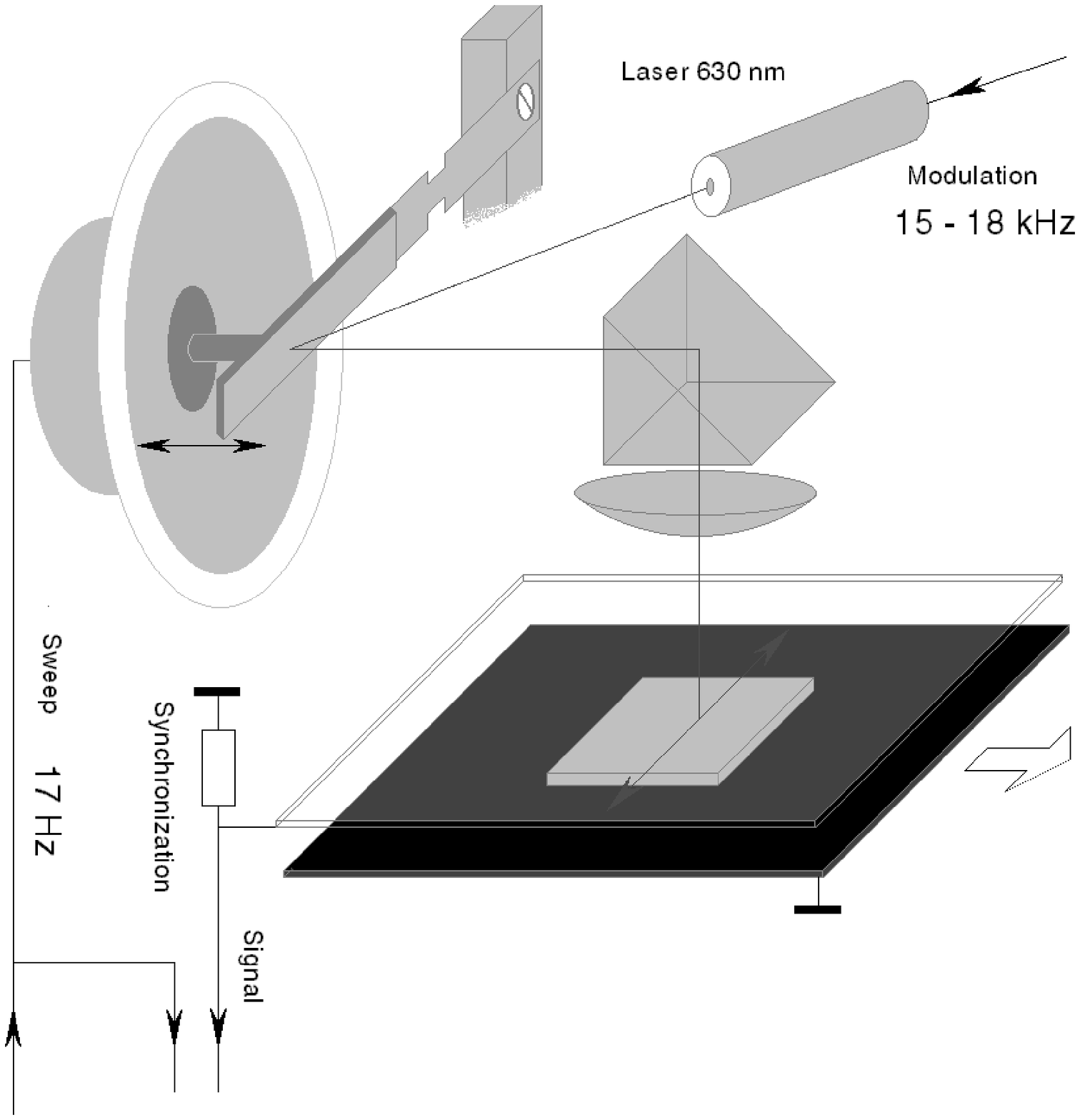}
\caption{
Schematic of the set-up for observation of the photo-emf relief
}
\end{figure}
\newpage
\begin{figure}
\epsfxsize=500pt
\epsffile{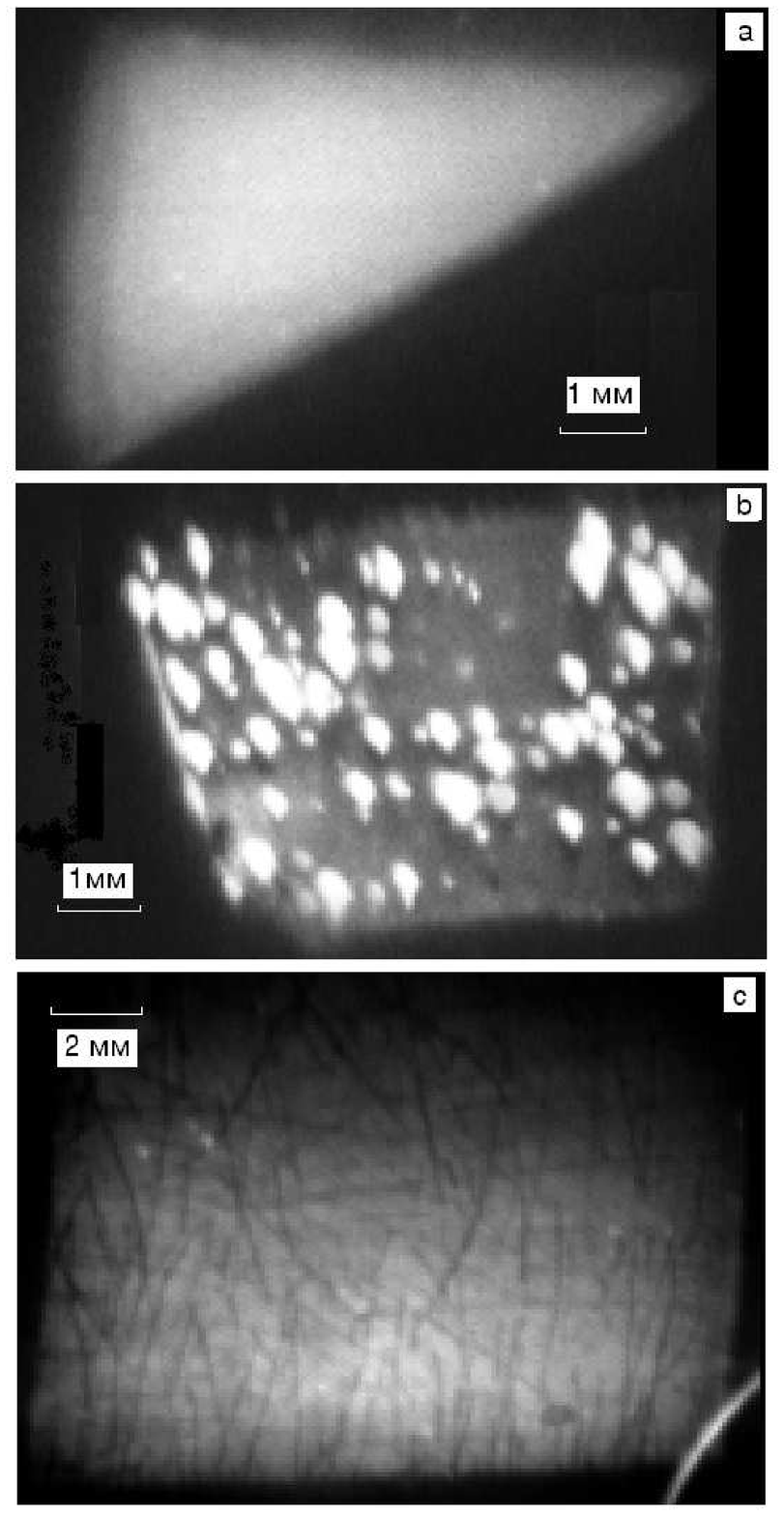}
\caption{
Examples of the relief of the photo-emf of a pure GaAs substrate (a), a planar
waveguide  (b), and a quantum well structure (c) grown using the MBE technique.
}
\end{figure}

\end{document}